\documentclass[sigconf,screen]{acmart}
\AtBeginDocument{%
  }

\setcopyright{none}
\renewcommand\footnotetextcopyrightpermission[1]{}

\begin{document}

\title{Teaching AI Interactively: An Experience Report in Higher Education}

\author{Jennifer M. Reddig}
\email{jreddig3@gatech.edu}
\orcid{0009-0000-6731-9401}
\affiliation{%
  \institution{Georgia Institute of Technology}
  \city{Atlanta}
  \state{Georgia}
  \country{USA}
}
\author{Scott Moon}
\email{scottqmoon@gmail.com}
\affiliation{%
  \institution{Independent Researcher}
  \city{Atlanta}
  \state{Georgia}
  \country{USA}
}

\author{Kaitlyn Crutcher}
\affiliation{%
  \institution{Georgia Institute of Technology}
  \city{Atlanta}
  \state{Georgia}
  \country{USA}
}
\email{kcrutcher3@gatech.edu}

\author{Christopher J. MacLellan}
\affiliation{%
  \institution{Georgia Institute of Technology}
  \city{Atlanta}
  \state{Georgia}
  \country{USA}
}
\email{cmaclell@gatech.edu}

\renewcommand{\shortauthors}{Reddig et al.}

\begin{abstract}

Introductory artificial intelligence (AI) courses present significant learning challenges due to abstract concepts, mathematical complexity, and students’ diverse technical backgrounds. 
This paper presents an experience report examining the redesign of in-class instructional time in a university-level Introduction to Artificial Intelligence course, inspired by CS Unplugged approaches.
We redesigned the summer offering, integrating embodied, unplugged simulations, collaborative programming labs, and structured reflection to provide students with a first-person perspective on AI decision-making. 
We maintained identical assignments, exams, and assessments as the traditional lecture-based offering. 
We found that students in the redesigned course reported higher attendance, stronger agreement that assessments measured their understanding, and greater overall course effectiveness, despite no significant differences in self-reported learning. 
Post-course interviews indicate that unplugged simulations and collaboration fostered a safe, supportive learning environment that increased engagement and confidence with AI concepts. 
These results highlight the importance of in-class instructional design in improving students’ learning experiences without compromising rigor.
\end{abstract}




\maketitle

\section{Introduction}



Artificial intelligence (AI) has become a core component of modern computing curricula, with introductory AI courses often serving as students’ first sustained 
encounter with probabilistic reasoning, search, learning algorithms, and ethical considerations surrounding intelligent systems. 
These courses play a critical role in shaping how students understand what AI is, how it works, and whether they view the field as accessible and relevant to their goals. 
However, AI concepts can be conceptually dense and mathematically demanding, creating barriers that can discourage engagement and undermine students’ confidence, particularly for those with less technical experience. 
As a result, how AI is taught has significant implications for student learning, persistence, and identity formation within the field.

Research of higher education AI courses most often focused on project-, problem-, and game-based learning, where students develop substantial AI artifacts to apply course concepts in realistic contexts \cite{1611941,vargas2020project,de2023using}.
However, these approaches primarily emphasize out-of-class project work and final deliverables \cite{ng2023review}.
In-class instruction is where students first encounter new ideas, ask questions, and decide whether AI feels accessible and is something they can see themselves pursuing further.
CS Unplugged is a promising approach for making computing concepts engaging and accessible.
Hands-on, unplugged activities can act as complementary strategies for supporting technical outcomes.
Understanding how these activities can support student engagement in higher-education AI courses is a relatively new field of research \cite{chai2025larger, park2026cs}.

This paper presents our experience redesigning and teaching a university introductory AI course to change how students engage with AI concepts during class time.
We introduced interactive unplugged activities with explicit guided plugged programming components. Using course evaluation data, we find that students reported higher attendance and greater perceived course effectiveness than the prior semester. Student interviews revealed how interactive activities and a tangible sense of progress contributed to these outcomes.

\section{Background}






Artificial intelligence is a challenging and technically demanding subject in higher education, due to its reliance on advanced mathematics, algorithmic reasoning, and programming skills. 
Introductory AI and machine learning courses typically require substantial prior preparation in areas such as discrete mathematics, calculus, linear algebra, and probability \cite{10893290}. 
Students consistently identify mathematical concepts as primary sources of difficulty, alongside challenges in translating theory into implementation, and describe these courses as overwhelming and time-intensive across levels of prior experience \cite{10.1145/3724363.3729107}. 
These challenges can negatively impact students’ self-efficacy and persistence, particularly when students lack confidence in their mathematical or technical abilities \cite{10.1145/3485062}. 






CS Unplugged is an established technique for physically engaging students in computing content, using technology-free, collaborative activities to make abstract computing concepts concrete through physical and embodied experiences \cite{bell2018cs}.
Unplugged activities can lower barriers to entry, increase motivation, and support conceptual understanding, especially for students with limited prior technical experience \cite{battal2021computer,huang2021critical}.
Recent research has extended unplugged approaches to K-12 AI education, demonstrating their use for introducing concepts such as agent behavior, semantic networks, and facial recognition through embodied analogies and interactive simulations \cite{10.1145/3626252.3630783, lindner2019unplugged,10.1145/3465074.3465078,lim2024unplugged}.
Applying unplugged activities to university CS education is a relatively recent venture. 
\citet{park2026cs} used unplugged activities in gateway CS courses and found that students reported greater confidence and fewer perceived challenges, and stronger feelings of instructor support when using these methods.
\citet{chai2025larger} used physically engaging activities to demonstrate machine learning techniques. Students reported increased engagement, retention, and found class time valuable.
Unplugged learning can help university students feel engaged during class and build their confidence when learning abstract and technically challenging AI content.

Importantly, unplugged activities are most effective when integrated into instructional sequences that transition students from conceptual exploration to guided programming practice \cite{ma2023developing, dai2025}. 
Concreteness fading \cite{bruner1974toward} explicitly links physical, embodied examples to fully generalizable, abstract concepts, and has been shown to support better conceptual understanding than just the physical activity or keeping concepts abstract in mathematics and science domains \cite{fyfe2014concreteness}.
Within computing education, unplugged activities designed to mirror subsequent programming tasks have been shown to support the transition from concrete to abstract \cite{kerr2025unplugged}. 
Embodied engagement during the unplugged task can play a role in how students reason and reduce the cognitive demand of the plugged task \cite{kerr2026children}.
Concreteness fading is most effective when at least one intermediate stage explicitly links the physical and abstract representations \cite{trory2024pirate, trory2026concreteness}.
In this work, we draw on these principles, using unplugged and hands-on activities to support engagement while explicitly bridging embodied understanding to code-based implementation. 
We redesigned an introductory AI course around a systematic unplugged-to-plugged instructional cycle.

\section{What We Did}





Our exploratory pilot focused on integrating embodied, unplugged activities into an existing introductory artificial intelligence course in a higher education setting. The course, Introduction to Artificial Intelligence, is a general survey course of AI
techniques and has four main topics: Search, Reasoning with Uncertainty, Probabilistic Reasoning through Time, and Machine Learning. The course is usually taught in the Spring semester, lasts 16 weeks, and has an enrollment of approximately 300 students. Our intervention took place during a 12-week Summer offering, which enrolled approximately 50 students. Despite the inherent differences between these two offerings, the ratio between the instructional team members and the number of students remained the same, and we kept the same projects, assessments, and general structure of the course the same. To the extent possible, we changed only the in-class time, from traditional lecture-based instruction, to an active learning loop integrating unplugged activities and transitioning to ``plugged'' activities. 
\footnote{All instructional resources are available online: https://github.com/jmreddig/Intro-to-AI-Teaching-Resources}


The Spring offering is taught entirely through lecture in large halls, with instructor and TA drop-in office hours available for one-on-one or small group remediation.
During the summer semester, we modified the in-class lecture time to focus on active and interactive instructional methods.
To introduce each topic, students participated in an unplugged simulation like the ones described by \citet{reddig2026aiunplugged}.
The simulation served as an accessible entry point to AI decision-making, and the mathematical formalization.
After a class discussion of the activity and a brief introduction of the formal algorithm, students then constructed the simulation in a Jupyter notebook.
The notebook activity provided scaffolding for the Python programming skills needed for the summative coding projects.
For more in-depth exploration of the intricacies of each algorithm, the instructional team also developed collaborative coding labs that allowed students to experiment with alternative parameters and problem settings.
At the end of each class session, students completed an exit ticket in the form of a One Minute Paper (OMP) \cite{stead2005review}.
Students were asked to describe what they now know about the day's topic, how what they did during class helped them learn each topic, and what questions they have now (points of confusion, clarification requests, musings about real-world applications, connections to other topics, etc).
The instructional team used the OMP submissions to revise the next session's lesson plan to match the student's actual progression through course content.


\begin{figure}
    \centering
    \includegraphics[width=1\linewidth]{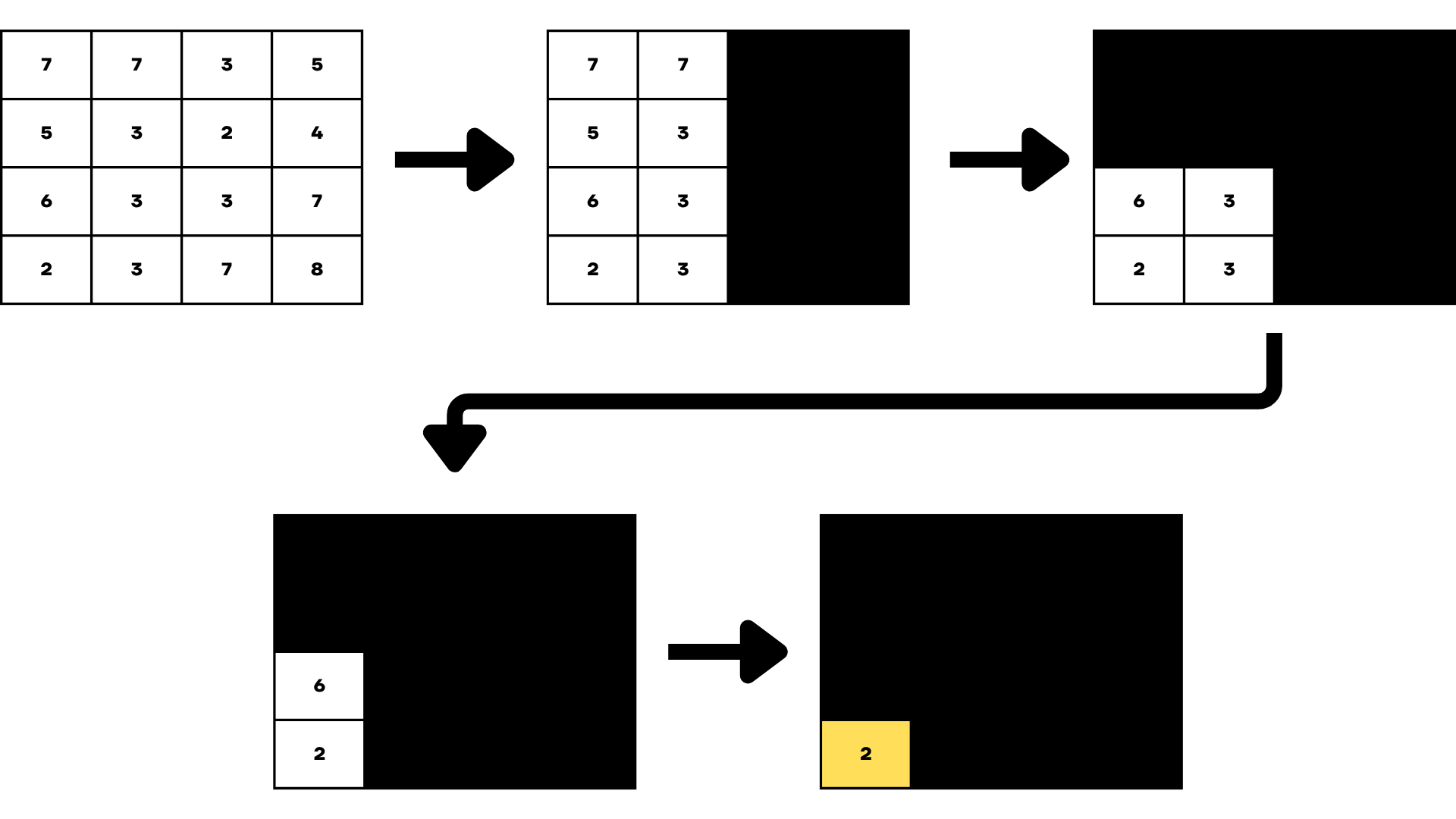}
    \caption{In Grid Knockout, players take turns eliminating half of the grid until only one number remains. `Max' players aim for a high number while `Min' players aim for a low number.}
    \label{fig:gridknockout}
\end{figure}

As an example, during the unit on adversarial search in a game context, we used an activity called Grid Knockout (see Figure~\ref{fig:gridknockout}). 
The game board consists of a grid of numbers. 
One player, the `Max' player, can eliminate either the top or bottom half of the grid during their turn. 
The `Min' player can eliminate either the left or right half. Players take turns until there is only one square left. 
The `Max' player would like a high number; the `Min' player would prefer a low number. 
We distributed several 8 by 8 and 4 by 4 grids to pairs of students for them to play during the concrete enacting phase \cite{suh2020we}. 

The advantage of using Grid Knockout is that the game is small enough to explore the game tree by hand.
Even seemingly simple examples like tic-tac-toe can quickly become unwieldy.
The game also has no real available strategies beyond looking ahead and mentally performing something like minimax mentally.
Games like tic-tac-toe may have distracting mental shortcuts like, ``Pick the center square.''
While Grid Knockout is easily visualized as a tree, it still requires a small step to translate it to that abstraction.
Despite it's simplicity, it is still relatively enjoyable to play.

After the activity, there was a discussion on strategies and the experience of playing.
For the smaller grids, students described looking ahead to the end of the game and working backwards, thinking about comparing pairs of numbers.
For larger grids, they generally chose based on perceived concentration of high or low numbers at a glance, which was useful to call back to later when discussing heuristic evaluation of general game states.
After the discussion, we used their growing intuition of looking ahead to introduce the minimax algorithm in pseudo-code and demonstrate a pen-and-paper example in a short lecture.
Students then attempted an example with their game partner.

Next, we provided students with a Jupyter notebook with a ready-made Grid Knockout game object and play/test environment.
Using the pseudo-code, student's coded their own abstract minimax algorithm implementation. 
After letting them try on their own, we modeled how to code the conditional branch where the Max player is to move.
They were then able to code the other branch with the Min player to move, flipping signs and comparisons as needed.
The students were able to run the example games through their own code, as well as generate and test new grids.

We then began a discussion on how these ideas might apply to real games.
At this point, our students were keenly aware of how small Grid Knockout is compared to chess or go.
We introduced the idea of a maximum depth and heuristics to avoid an exploding tree, and cued up alpha-beta pruning by asking how we could avoid searching every possible branching path, knowing that our opponent will not make bad moves.
After another round of pen-and-paper examples, this time using alpha-beta pruning, students returned to their Jupyter notebooks to add alpha-beta pruning.

Finally, they completed a brief exit ticket in the form of a One Minute Paper.
We asked the following questions:
\begin{itemize}
    \item What was the most important thing you took away from today’s activity?
    \item In your own words, tell me what you know about adversarial search.
    \item How did this activity help you understand the topic?
    \item What is still confusing or unclear about the topic?
\end{itemize}
We used that information either for targeted interventions addressing misconceptions or review and clarification at the beginning of the next class period.

For other topics, the general outline remained the same.
We changed the activity, using other unplugged simulation activities like those in \citet{reddig2026aiunplugged}, and wrote Jupyter notebooks to build on the particulars of that activity.
We alternated between discussion, short lecture, and active coding until the end of class and the OMP.
This unplugged-to-plugged active learning loop was our main change to the course.
Again, the assessments, projects, and content order remained the same.

After completion of the course, we invited students to complete the anonymous institution end of course survey so we could compare the previous semester to the redesigned course.
The survey asked students to self-report their preparedness to take the course at the start, how well the assignments measured their knowledge, the percentage of assignments completed, and the percentage of classes attended.
It also asked students to rate the amount they learned from the course and the overall course effectiveness.
Each survey item is on a 5-point Likert scale, except the two percentage questions which are on a 6-point Likert scale.
Response rate from both the Spring and Summer semesters was \~25\%.

In addition to the survey, we invited the students to participate in a retrospective semi-structured interview of their experience learning artificial intelligence. 
We conducted post-course interviews after all graded materials were submitted and final scores released so that student responses would not be influenced by their desire to achieve a particular grade. 
The interview process was reviewed and approved by the authors’ Institutional Review Board. 
All participants provided informed consent prior to participation.
We conducted interviews with eleven students (4 female, 7 male), asking them to recall and reflect on their time in class. 
We conducted the interviews via Zoom and recorded after obtaining verbal consent.
After transcribing the interviews manually from the recording, we cleaned and anonymized each transcript before analysis and subsequently deleted the recordings.

\section{What We Found}

\begin{table}[]
\caption{Ordinal logistic regression coefficients estimating the effect of semester on student survey ratings of course experiences.}\label{tab:survey_items}
\begin{tabular}{|c|c|c|c|c|}
\hline
\textbf{Survey Item}                          & \textbf{$\beta$} & \textbf{SE} & \textbf{z}   & \textbf{p} \\ \hline\hline
Preparedness & 0.37   & 0.57      & 0.66 & 0.510   \\ \hline
Homework Completed     & 0.68   & 1.09      & 0.62 & 0.533   \\ \hline
Amount Learned             & 0.77   & 0.56      & 1.39 & 0.164   \\ \hline
Class Attended       & 1.92   & 0.58      & 3.31  & 0.001*  \\ \hline
Assignments       & 1.44  & 0.60      & 2.40 & 0.016*  \\ \hline
Effectiveness         & 1.22   & 0.62      & 1.95  & 0.050*  \\ \hline
\end{tabular}
\end{table}

\begin{figure}
    \centering
    \includegraphics[width=0.8\linewidth]{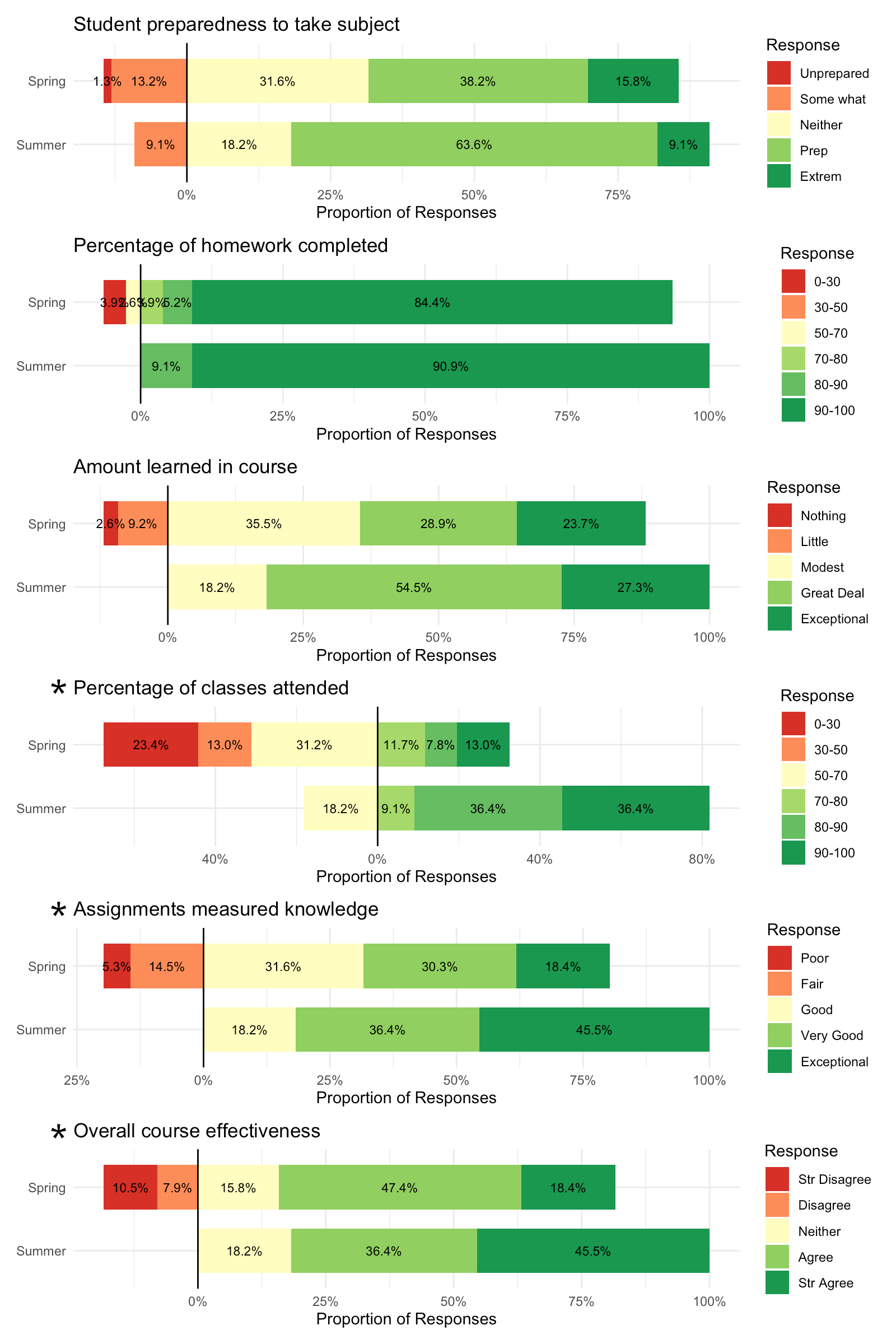}
    \caption{Stacked proportional bar charts showing student responses to survey items comparing Spring and Summer semesters. Statistically significant responses are marked with an asterisk (*).}
    \label{fig:likert}
    \Description{
    Stacked bar charts show the percentage distribution of student responses for six survey measures: preparedness to take the subject, percentage of homework completed, amount learned in the course, percentage of classes attended, perceptions that assignments measured knowledge, and overall course effectiveness. Responses are grouped by semester (Spring and Summer) and displayed as proportions within each response category. Overall, negative responses were substantially reduced in the Summer offering, while positive responses (“Prepared,” “Great Deal,” “Very Good,” “Exceptional,” “Agree,” and “Strongly Agree”) were more prevalent across all measures.
    Percentages of classes attended, assignments measured knowledge, and overall course effectiveness are labeled as statistically significant differences.
    }
\end{figure}

We conducted an ordinal regression to examine whether the different instructional strategies predicted student ratings on the end-of-semester course evaluation survey. 
For each survey item, we fit an ordinal logistic regression model with semester as the predictor variable and the survey response as the outcome.
Differences in student scores are represented in Figure~\ref{fig:likert} and regression coefficients for all items are listed in Table~\ref{tab:survey_items}.  

Three of the six survey items responses had statistically significant differences between the two semesters. 
When asked what percentage of classes they were physically present for, student answers were significantly different, Wald $z = 3.3, p < 0.001$. 
Students in the summer session had 6.8 times higher odds of reporting higher attendance than students in the spring session $(\beta = 1.92, SE = 0.58)$.
We also see significant differences in how students perceived the assignments, Wald $z = 2.40, p < .016$. 
Though students in both semesters completed the same projects and exams, students in the summer semester were approximately four times more likely to favorably report degree to which the exams, homework, and other assignments measured their knowledge and understanding $(\beta = 1.44, SE = 0.60)$.
Finally, there is also a significant difference in the student ratings for overall course effectiveness, Wald $z = 1.96, p < .05$. 
Students in the summer session had 3.4 times higher odds of agreeing that the course was overall effective $(\beta = 1.22, SE = 0.62)$.
The remaining three survey questions did not differ significantly.

\begin{figure}
    \centering
    \includegraphics[width=1\linewidth]{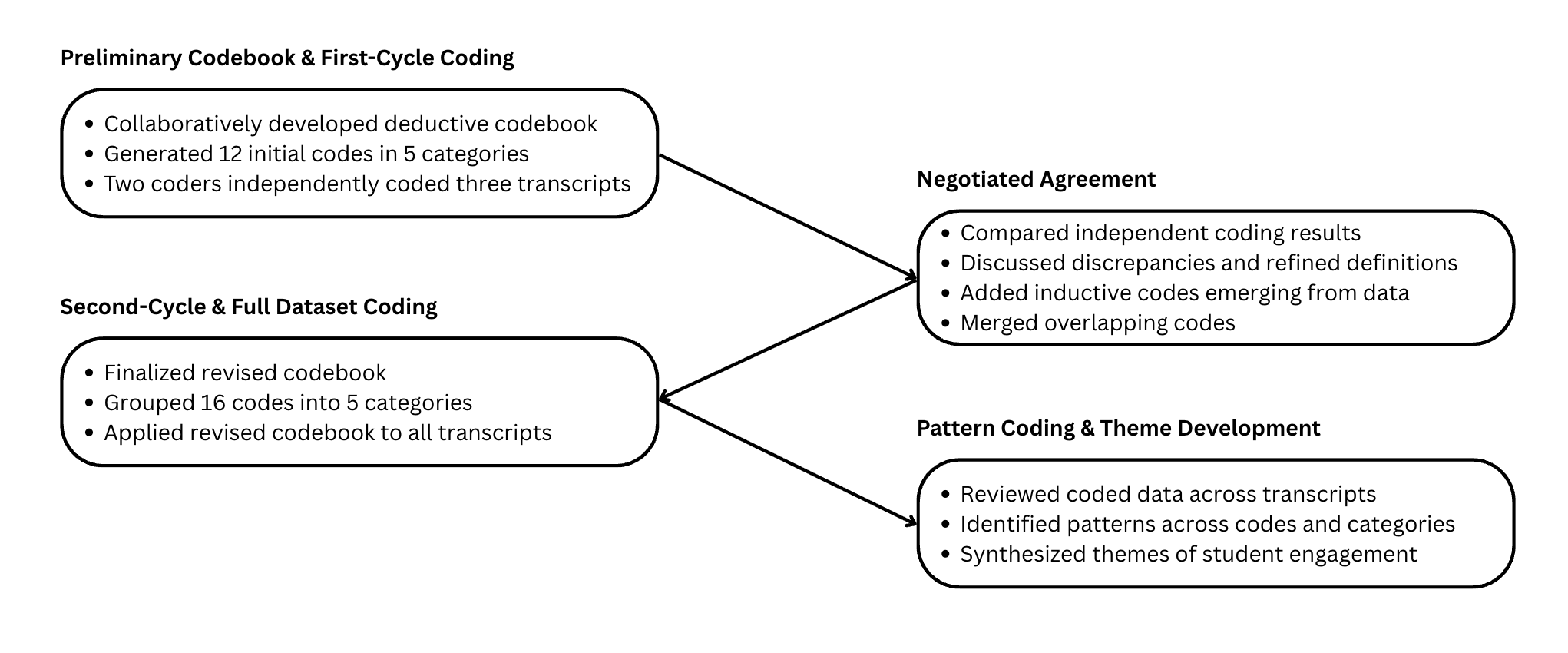}
    \caption{Qualitative Data Analysis Process}
    \label{fig:qual_process}
\end{figure}

We used a two-cycle coding process to analyze student interview transcripts to better understand the student experience, detailed in Figure~\ref{fig:qual_process} \cite{saldana2021coding}.
In the first cycle, the research team built a preliminary code-book.
Two independent coders then coded 3 of the 11 transcripts individually with the preliminary deductive codebook.
After reviewing the results, several additional codes emerged inductively about the classroom environment..
The revised codebook included 16 codes grouped into 5 categories.
With the updated codebook, we coded the remaining eight transcripts.
We then reviewed the coded data and applied an inductive process to identify patterns and themes across codes and categories \cite{bingham2021deductive}. 
We identified four themes within the data that students attributed to their engagement with the course described below.

\subsubsection{Actively Constructing Knowledge}

Students enjoyed the ways the class sessions helped them engage with the content in a hands-on manner.
They appreciated the unplugged simulation activities, and often described their favorite activity as an example of a time they felt engaged. 
The activities allowed students to visualize the algorithm physically and play through small details in a hands-on manner (\textit{``The in class activities really broke concepts down to their core components and allowed us to }be\textit{ the components and perform actions, the way that they would be performed by an AI agent.''}). 
While learning the mathematics behind an AI algorithm can be challenging, the in-class instruction helped break down the challenge into manageable pieces.
The coding labs helped students transition from conceptual understanding to practical implementation, in a low-stakes way that prepared them to complete the graded programming assignments.
Students described learning as an active process of building understanding through doing rather than passively receiving information.





\subsubsection{Safe, Supportive Learning Environment}
The unplugged activities provided a step-by-step breakdown of each algorithm which clarified conceptual misunderstandings and helped students understand what their code implementation should do line-by-line.
Collaboration during the in-class activities helped students build connections, make friends, and find someone to share with and learn together (\textit{``I was able to have chances to talk to the people around me and they were able to help me understand concepts I didn't understand and I was able to share things with them''}).
The constant presence of the instructor and TAs made it very easy to receive timely, personal support.
Students also found attending class enjoyable, perhaps because the classroom culture welcomed questions and acknowledged confusion as part of the learning process (\textit{``I actually enjoyed the content because I felt like I was doing something fun with each activity and I also felt it was a more safe environment for me to be able to ask questions.''}).
This safe and supportive atmosphere helped students engage deeply with complex content while feeling comfortable asking for help and learning through uncertainty.





\subsubsection{Sense of Progress}
Students felt that attending class was worthwhile, because they were going to get something valuable and meaningful each day (\textit{``Seeing something come to life and with an actual lab or some type of activity that made you feel you weren't just learning it, you were using it and it had a purpose... It made me feel the time coming to class was worth it, which is the only reason I attended so many classes versus staying at home.''}).
Students also saw content from other courses, like Data Structures and Probability, applied in new ways.
Connecting AI concepts to other classes made both courses feel more practical.
This cumulative sense of progress reinforced the value of attending class and motivated sustained engagement throughout the semester.

\subsubsection{Increased Confidence}
Students tackled many challenging concepts: formulas with lots of variables, plenty of math operations, learning to code in Python for the first time, and demystifying black-box AI methods like neural nets.
Breaking each concept down into manageable pieces through the unplugged simulations and guided programming labs made the content accessible with an easy entry point, so students could see that AI methods are attainable (\textit{``And I guess like how, I don't want to say how simple it is because it really isn't that simple, but it's a lot less scary than I thought it was.''}).
Discussing, using, and coding AI every day in class helped students build their confidence as an AI engineer, especially those that had attempted and dropped this class in a prior semester.
Students attitudes shifted, from thinking AI is inaccessible and complicated to visualizing themselves as AI engineers one day (\textit{``[I used to say] `I don't think I would ever become a engineer in this. I don't think I have the skill set. I don't think this is ever doable'. And it got pulled down the earth really quickly and I realized that I want to do it now.''}).














The safe, supportive classroom environment and tangible sense of progress students described likely indicate why attendance was significantly higher in the redesigned course.
Students found value in showing up because each session offered something they could not replicate on their own. 
Of particular interest is the finding that students rated assignments as better measures of their knowledge despite the assignments being identical across semesters.
Students remarked that the in-class coding labs helped them prepare to tackle the more involved summative coding projects.
Paired with the interviews, we speculate that because students were more engaged and felt the challenge was manageable, that they were more willing to see the assignments as an accurate reflection of their progress. 

Students also reported feeling more comfortable discussing and explaining AI methods.
This takeaway was especially important for students who reported limited prior programming experience, students who had not taken a CS class before, and students who had previously withdrawn from the course.
\citet{hogan2023cs0} found that students in introductory CS courses that are concerned about their lack of preparation and experience had the greatest gains in confidence.
Students who initially doubted their ability to succeed and learn AI were able to not only acquire new skills, but improve their confidence.
They changed their self-image into someone who could have meaningful conversations with AI professionals, even expressing a desire to continue their AI studies beyond what they had previously imagined
(``\textit{I was like I'm not doing [an AI concentration] anymore after [I dropped the course in the spring], but I'm glad I stuck with it because I am 100\% more comfortable with AI and I really want to continue it now.}'').

\section{Discussion}

\subsection{Implications for Instructional Design}

This course redesign demonstrates the importance of making the time students spend in class meaningful. 
Students need to find value in attending lecture sessions beyond earning an attendance grade or passively receiving content to keep them coming to every session.
Using a variety of active learning techniques and centering students during discussions helps keep students engaged through the session and the semester
(\textit{``It wasn't just a lecture where I just had to sit and listen the entire time..., even though the class lecture was longer, almost two hours, I was able to just focus the entire two hours, which I never have been before in like other classes.''}).
Even though assignments, exams, and out-of-class expectations were unchanged, students were significantly more likely to attend class when in-class activities provided tangible progress and immediate payoff.
This pattern aligns with expectancy–value theory, which suggests that students are more motivated to engage in tasks when they perceive them as both valuable and directly connected to their success \cite{wigfield1994expectancy, wigfield2021achievement}.
Attendance should not be motivated by course policies or incentives, but of whether instructional time offers learning experiences that students perceive as distinct from what they could achieve on their own.

It also appears that tightly coupling in-class and out-of-class activities appears to amplify the meaning of both.
Unplugged simulations, collaborative coding labs, and structured reflection were explicitly connected to the summative programming assignments students completed outside of class. 
As a result, students reported that assignments better measured their understanding than the delivery of the exact same assignments in the prior semester.
The alignment helped students see coursework as a learning trajectory rather than a collection of disconnected tasks, reinforcing a sense of progress and purpose throughout the semester
For instructors looking to redesign their course, consider how to bring your in-class and out-of-class activities closer together to strengthen the instructional through-line
(``\textit{I think when the activity was tightly coupled with the topic was when it was the most engaging.}'').
Changing how students learn can be just as important as what they produce.

\subsection{Limitations}

This experience report reflects a single implementation at one institution, and several contextual factors shaped the experience in ways that other practitioners should keep in mind.
The Summer offering enrolled approximately 50 students over 12 weeks, compared to approximately 300 students over a 16-week Spring semester. 
These structural differences may have influenced student engagement, classroom dynamics, and help-seeking behaviors independent of the instructional redesign.
Differences in instructor experience, teaching style, or rapport with students may have contributed to the observed outcomes.
Although the course materials, projects, and content were held constant across semesters, instructor effects cannot be fully separated from the impact of the instructional redesign.

Our reflections draw on course evaluations, retrospective interviews, and observational notes. 
The interviews involved eleven students, which provided useful qualitative depth but limited breadth. 
As with any self-reported data, responses may reflect recall bias or social desirability effects \cite{podsakoff2003common}. 
The implementation also benefited from strong teaching assistant involvement during class sessions. 
This was an important enabler of the approach as described, and instructors in settings with limited TA support should factor this into any adaptation.

\subsection{Scaling Up}


One important factor in the redesigned course was the active presence of teaching assistants during class sessions. 
Teaching assistants provided real-time support, answered questions, and helped facilitate collaborative activities. 
We believe that with greater involvement from teaching assistants, the active and student-centered instruction described in this paper can scale to the larger Spring sessions.
Prior work \cite{essick2016scaling, silva2022innovating} provides an example for how TAs can be trained and utilized to scale-up collaborative learning in large classes.
With appropriate training and clearly defined facilitation roles, TAs can support subgroups of students in the collaborative, unplugged simulations and surface group misconceptions and confusion areas to the instructor to address more broadly.
Coordinating classroom management between TAs and the instructor would allow each student to receive individualized attention from at least one member of the instructional team and allow the instructor to concentrate in-class support where most needed.
Many of the unplugged simulations and collaborative programming labs were intentionally designed to be flexible with the number of students in each group. 
In a large-enrollment setting, modifying the group size and using teaching assistants to organize and facilitate peer collaboration can help extend the instructional team's reach to make learning feel personal and individual.

The use of frequent formative feedback, such as One-Minute Papers, may become even more valuable at scale.
Aggregated student reflections can provide instructors with rapid insight into common misconceptions and guide targeted adjustments to subsequent class sessions.
Reshaping feedback mechanisms to better suit aggregation, like surveys, polls, and quizzes, would make it easier for instructors to respond to the student experience.
Alternatively, using TAs to collect, aggregate, and interpret OMPs would let the instructional team receive qualitative responses about student progress.

\subsection{Lessons Learned}

As a pilot, this course redesign provides insight into how future iterations could be strengthened.

\subsubsection{Tighten simulation to formalization}

The most successful unplugged simulations closely mirrored the structure of the underlying mathematics. 
When the mechanics of the simulation reflected the actual update equations or algorithmic steps, students were able to transition more smoothly to formal notation and code implementation. 
In contrast, when a simulation abstracted or simplified the mathematics too heavily, students sometimes struggled to map their experiential understanding onto the symbolic formulation. 
Future iterations will prioritize tighter structural alignment between simulation design and the targeted algorithm. 
The closer the correspondence between the embodied experience and the mathematical model, the more seamless the transition to formal derivation and programming becomes.

\subsubsection{Clarify the transition from experience to formalization.}

While unplugged simulations provided accessible entry points into complex AI algorithms, students benefited most when the transition to formal mathematics was made explicit. 
Making this bridge visible helps students better articulate how interactive activities connect to theory.
Future iterations will place greater emphasis on systematically annotating the mapping between the simulation and the algorithm. 
Rather than assuming students will infer how the physical experience connects to symbolic notation, instructional explanations will more deliberately trace how each step of the simulation reflects the structure of the algorithm.

\subsubsection{Strengthen explicit connections to assessments.}

Students benefited from additional structured time to apply key frameworks and tools, like \texttt{pgmpy} or \texttt{PyTorch}, in ways that closely resembles project tasks.
Giving students chances to construct derivations, interpret algorithm behavior, debug model implementations, and reason about parameter choices during class can reduce the gap between exploratory activity and high-stakes assessment. 
When in-class practice mirrors the cognitive and technical demands of exams and projects, students can build fluency and confidence in applying concepts independently.
Future offerings will therefore incorporate targeted rehearsal problems and scaffolded coding challenges that reflect the structure and expectations of summative assessments.

\section{Conclusion}


This pilot provides promising initial evidence that redesigning instructional time can meaningfully improve students’ experiences in introductory AI courses.
By replacing primarily lecture-based class sessions with interactive, unplugged, and collaborative activities, we observed higher student attendance, stronger perceptions that assessments measured understanding, and greater overall course effectiveness.
Embodied activities, a close alignment between in-class and out-of-class work, and proactive instructional support built engagement, confidence, and a sense of progress with challenging AI concepts in our students.
These results suggest that in-class instructional design plays a critical role in shaping how students interpret difficulty, value coursework, and see themselves as capable learners in AI.
Future work should examine whether these effects hold at scale and across more diverse institutional contexts.





\bibliographystyle{ACM-Reference-Format}
\bibliography{sample-base}










\end{document}